\documentclass[journal=ancac3,manuscript=article]{achemso}

\usepackage[version=3]{mhchem} % Formula subscripts using \ce{}
\usepackage[T1]{fontenc}       % Use modern font encodings
\usepackage{mathtools}

\author{Stefano~Dal~Forno}
\email{tenobaldi@gmail.com}
\affiliation{Department of Physics, Imperial College London, London SW7 2AZ, United Kingdom.}
\author{Luigi~Ranno}
\affiliation{Department of Materials, Imperial College London, London SW7 2AZ, United Kingdom.}
\author{Johannes~Lischner}
\email{j.lischner@imperial.ac.uk}
\affiliation{Department of Physics and Department of Materials, and the Thomas Young Centre for Theory and Simulation of Materials, Imperial College London, London SW7 2AZ, United Kingdom}

\title[Hot-carriers in plasmonic nanoparticles]{Material, size and environment dependence of plasmon-induced hot carriers in metallic nanoparticles}


\begin{document}

\begin{tocentry}
\includegraphics{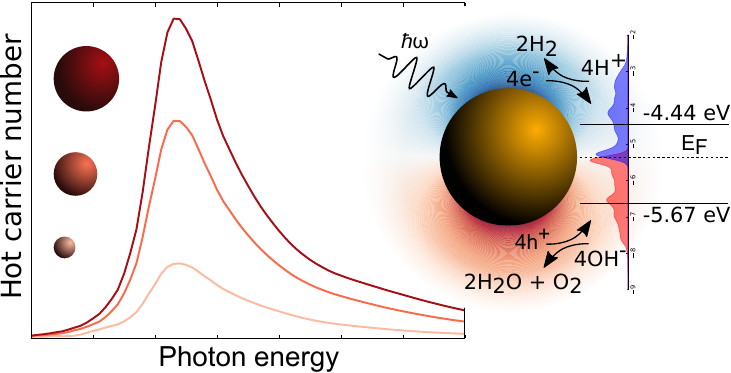}
\end{tocentry}

\begin{abstract}
Harnessing hot electrons and holes resulting from the decay of localized surface plasmons in nanomaterials has recently led to new devices for photovoltaics, photocatalysis and optoelectronics. Properties of hot carriers are highly tunable and in this work we investigate their dependence on the material, size and environment of spherical metallic nanoparticles. In particular, we carry out theoretical calculations of hot carrier generation rates and energy distributions for six different plasmonic materials (Na, K, Al, Cu, Ag and Au). The plasmon decay into hot electron-hole pairs is described via Fermi's Golden Rule using the quasistatic approximation for optical properties and a spherical well potential for the electronic structure. We present results for nanoparticles with diameters up to 40 nm, which are embedded in different dielectric media. We find that small nanoparticles with diameters of 16 nm or less in media with large dielectric constants produce most hot carriers. Among the different materials, Na, K and Au generate most hot carriers. We also investigate hot-carrier induced water splitting and find that simple-metal nanoparticles are useful for initiating the hydrogen evolution reaction, while transition-metal nanoparticles produce dominantly holes for the oxygen evolution reaction.
\end{abstract}

Energetic or ``hot'' electrons and holes produced by the decay of localized surface plasmons (LSP) in metallic nanostructures have recently generated much excitement. They can be harnessed in optoelectronic devices, such as photodetectors, or for solar energy conversion, i.e. in photocatalytic or photovoltaic devices \cite{hotelectrons1,molecules1,watersplitting1,watersplitting2,watersplitting3,fuelproduction1,photocatalysis1,photocatalysis2}. For example, \citeauthor{mukherjee} observed that plasmon-induced hot electrons can trigger H$_2$ dissociation reactions on the surface of gold nanoparticles \cite{mukherjee}. An important advantage of nanoplasmonic devices compared to traditional systems is their tunability: their optical and electronic properties depend sensitively on the nanoparticle size and shape, but also on the nanoparticle material and its environment \cite{dionne,nordlander2,size_env_dependence1,govorov,fang,tune1}. 

To guide experimental progress and identify nano-devices with favorable hot-carrier properties, a detailed theoretical understanding of the physico-chemical processes that govern hot-carrier generation is needed. However, developing such a theory is challenging because of the large size of experimentally relevant nanoparticles. Atomistic \emph{ab initio} calculations are currently only feasible for metallic clusters and very small nanoparticles \cite{prineha_review,watersplitting3,silver_cluster_plasmon}. To model properties of experimentally relevant nanoparticles with radii of 10 nm or more, two different strategies have been employed. In many calculations, simplified models for the electronic structure of the nanoparticle are used, such as jellium models or non-interacting electron models \cite{ekardt,nordlander1,jellium_nanorods,prezhdo}. For example, \citeauthor{nordlander1} employed a spherical well model to simulate hot-carrier generation in silver nanoparticles with diameters up to 25 nm. Such approaches are accurate for nanostructures made of simple metals (i.e. metals with conduction electrons in s- or p-states), but are less reliable for transition-metal nanostructures where d-electrons can play an important role \cite{bernardi,ravishankar2}. 

Another class of approaches is based on the assumption that the electronic structure of the nanomaterial is similar to the electronic structure of the bulk material, which can be obtained with \emph{ab initio} approaches. These approaches are particularly useful for materials with shallow d-states, such as gold or copper \cite{bernardi,prineha,ravishankar1,ravishankar2}. However, a complete description of hot-carrier generation that captures both nanoparticle size effects and d-bands is still missing.  

In this paper, we do not attempt to develop such a complete theory of hot-carrier generation. Instead, we use a simplified description of the nanoparticle electronic structure to explore the dependence of hot-carrier properties on the nanoparticle material, size and environment. In particular, we carry out calculations for spherical nanoparticles with diameters up to 40 nm and investigate six different plasmonic metals: the simple-metals sodium (Na), potassium (K) and aluminium (Al) and the transition-metals gold (Au), silver (Ag) and copper (Cu). For some systems, such as simple-metal nanoparticles or transition-metal nanoparticles in an environment with a large dielectric constant, we expect that our simplified electronic structure model gives accurate results. For transition-metal nanoparticles in an environment with weak screening, our calculations are less accurate because of the lack of d-states in the model, but should provide a useful \emph{lower bound} on hot-carrier generation rates. 

The paper is structured as follows: we first explain in detail our computational approach for calculating hot-carrier generation rates emphasizing the importance of an accurate description of carrier lifetimes. We then present our results for the optical and hot-carrier properties of metallic nanoparticles, discuss consequences for hot-carrier induced solar water splitting and offer conclusions.

%%%%%%%%%%%%%%%%%%%%%%%%%%%%%%%%%%%%%%%%%%%%%%%%%%%%%%%%%%%%%%%%%%%%%%%%%%%%%%%%%%%%%%%%%%%%%%%%%%%%%%%%%%%%%%%

\section{Result and discussion}

\textbf{Description of the model.} For a spherical nanoparticle of radius $R$ and volume $V=4 \pi R^3/3$, illuminated by light of frequency $\omega$,
we use Fermi's golden rule to express the number of hot electrons with energy $E$ generated per unit time, volume and energy as
\begin{equation}
N_e(E,\omega)=\frac{2}{V} \sum_{if}\Gamma_{if}(\omega)\delta(E-E_f),
\label{eq:distribution}
\end{equation}
where the prefactor of 2 takes spin into account and $\Gamma_{if}$ is the probability
rate of exciting an electron from state $\Psi_i$ to state $\Psi_f$ (with corresponding energies $E_i$ and $E_f$) given by 
\begin{equation}
\Gamma_{if}(\omega)=\frac{2 \pi}{\hbar}|\langle \Psi_f|\Phi_{tot}(\omega)|\Psi_i \rangle |^2 \, \rho_{if}.
\label{eq:goldenrule}
\end{equation}
In this equation, $\Phi_{tot}$ denotes the total potential including both the external perturbation by the light field and the induced response of the nanoparticle, and $\rho_{if}$ describes the density of available states via\cite{cohen}
\begin{equation}
\rho_{if}(\omega)=\frac{\gamma_{if}}{\pi} \frac{1}{(\hbar\omega-[E_f-E_i])^2+\gamma_{if}^2}
+ \frac{\gamma_{if}}{\pi} \frac{1}{(\hbar\omega+[E_f-E_i])^2+\gamma_{if}^2},
\label{eq:finalstates}
\end{equation}
where the first term captures resonant transitions, while the second term describes anti-resonant transitions. The relative importance of these two processes is controlled by the linewidth of the transition $\gamma_{if}$ which plays an important role for the distribution of hot carriers\cite{nordlander1}. To calculate the distribution of hot holes $N_h(E,\omega)$, $E_f$ on the right hand side of Eq.~\ref{eq:distribution} has to be replaced by $E_i$.

To calculate $\gamma_{if}$ we use Matthiessen's rule and partition the linewidth into contributions from electron-electron (el-el) and
electron-phonon (el-ph) interactions in the initial and final states according to
\begin{equation}
\gamma_{if} = \gamma_{el-el}^i + \gamma_{el-ph}^i + \gamma_{el-el}^f + \gamma_{el-ph}^f.
\label{eq:matthiensen}
\end{equation}
The electron-phonon contributions to the linewidth are calculated with the Debye model \cite{sklyadneva,hofmann} according to
\begin{equation}
\gamma_{el-ph}^j = \frac{2 \pi \lambda \hbar \omega_D}{3},
\label{eq:debye}
\end{equation}
where $\omega_D$ denotes the Debye frequency corresponding to a Debye temperature $\Theta_D=\hbar \omega_D/k_B$ and $\lambda$ is the electron-phonon mass enhancment parameter.

For the electron-electron contributions to $\gamma_{if}$, Fermi liquid theory \cite{fermiliquid1,fermiliquid2} for a homogeneous electron gas of conduction electron density $\rho$ yields
\begin{equation}
\begin{split}
\gamma_{el-el}^j = \frac{m e^4 (E_j-E_F)^2}{64 \pi^3 \hbar^2 \epsilon_0^2 E_s^{3/2} E_F^{1/2}}
\left( \frac{2\sqrt{E_s E_F}}{4E_F + E_s} + \arctan \sqrt{\frac{4E_F}{E_s}} \right),
\end{split}
\label{eq:fermi_liquid}
\end{equation}
with $m$, $e$, $\epsilon_0$ and $E_F=\frac{1}{2}(3 \pi^2 \rho)^{2/3}$ denoting the electron mass, the electron unit charge,
the vacuum permittivity and the Fermi energy, respectively. Also, $E_S=\hbar^2 q_s^2/2m$ is the kinetic energy associated with the Thomas-Fermi
screening vector $q_s^2=\frac{e^2}{\epsilon_0}g(E_F)$, where $g(E_F)$ is the density of states at the Fermi level. Table~\ref{tab:parameters} summarizes all parameters that were used in the calculation of $\gamma_{if}$.

\begin{table}
\caption{Conduction electron densities $\rho$ (in nm$^{-3}$), work functions $\phi$ (in eV), Debye energies $k_B \Theta_D$ (in eV), electron-phonon coupling parameters $\lambda$ (dimensionless) and the resulting electron-phonon linewidths $\gamma_{el-ph}$ (in meV) of the selected metals.}
\begin{center}
\begin{tabular}{ccccccccc}
\cline{3-9}
&  & Na & K & Al & Cu & Ag & Au & Ref. \\
\hline
& $ \rho $  & 25.36 & 13.18 & 180.8      & 84.53      & 58.56      & 59.01      & \cite{handbook} \\
&$\phi$     & 2.36  & 2.29  & 4.20 (100) & 5.10 (100) & 4.64 (100) & 5.47 (100) & \cite{handbook} \\
&           &       &       & 4.06 (110) & 4.48 (110) & 4.52 (110) & 5.37 (110) & \\
&           &       &       & 4.26 (111) & 4.94 (111) & 4.74 (111) & 5.31 (111) & \\
&           &       &       &            & 4.53 (112) &            &            & \\

\hline
& $k_B \Theta_D$    & 0.013 & 0.009 & 0.034 & 0.027 & 0.019 & 0.015 & \cite{Ashcroft} \\
& $\lambda$         & 0.18 & 0.13 & 0.45 & 0.13 & 0.12 & 0.17 & \cite{Grimvall} \\
& $\gamma_{el-ph}$  & 4.9 & 2.3 & 32.0 & 7.4 & 4.7 & 5.2 & \\
\hline
\end{tabular}
\end{center}
\label{tab:parameters}
\end{table}

\begin{figure}[t]
\includegraphics{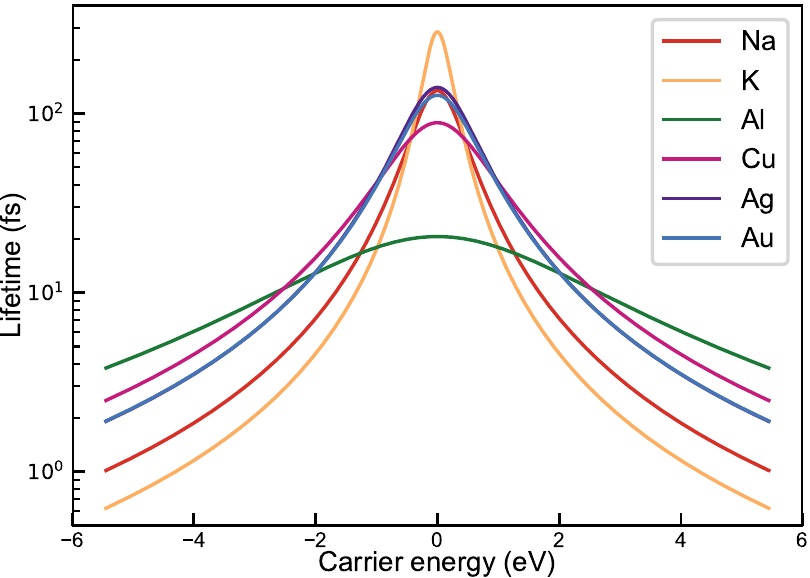}
\caption{Hot carrier lifetimes due to electron-phonon and electron-electron interactions for six different metals. The zero of energy is set to the Fermi level.
The lifetimes span a time range
between 1 and 300 femtoseconds.}
\label{fig:lifetimes}
\end{figure}

Figure \ref{fig:lifetimes} shows the calculated hot carrier lifetimes $\tau_{if}=\hbar/\gamma_{if}$ as a function of the hot carrier energies for the six selected metals. We observe lifetimes of up to several hundred femtoseconds
for states near the Fermi level and less than 10 femtoseconds at carrier energies higher than 2-3 eV.
These values are in good agreement with previous theoretical and experimental pump-probe results 
\cite{bernardi,nordlander2,sklyadneva,echenique,voisin,arbouet}. Note that the discrepancy with two-photon photoemission experiments, which often report lifetimes in the picosecond range\cite{klein,fermiliquid1,delfatti,govorov}, can be explained by transport effects\cite{aeschlimann,knoesel}.

The total potential in Eq. \ref{eq:goldenrule} is calculated within the quasi-static approximation\cite{plasmonics2} given by
\begin{equation}
\Phi_{tot}(\omega,\mathbf{r}) = -E_0 r \cos(\theta) +
E_0 \frac{\epsilon(\omega) - \epsilon_m}{\epsilon(\omega) + 2 \epsilon_m}
\begin{dcases}
r\cos(\theta), & r \le R, \\
 R^3 \frac{\cos(\theta)}{r^2}, & r > R,
\end{dcases}
\label{totalpotential}
\end{equation}
where the first term describes the perturbing light field and the second term captures the response of the
nanoparticle. Also, $E_0$ is the strength of the external electric
field which is set to ensure an illumination intensity of $1$~mW/$\mu$m$^2$ and 
$\epsilon(\omega)$ and $\epsilon_m$ are the dielectric functions of the bulk
material and the environment surrounding the nanoparticle, respectively. The bulk dielectric functions are calculated from experimental data for the refractive index $n(\omega)$ and the extinction index $\kappa(\omega)$\cite{handbook} via the standard formulas
\begin{align}
\epsilon(\omega)&=\epsilon_1(\omega)+i \epsilon_2(\omega), \\
\epsilon_1(\omega)&=n(\omega)^2-\kappa(\omega)^2, \\
\epsilon_2(\omega)&=2 n(\omega) \kappa(\omega),
\end{align}
where $\epsilon_1(\omega)$ and $\epsilon_2(\omega)$
denote the real and imaginary parts of the dielectric function, respectively.

The final ingredients needed to evaluate Eqs.~\ref{eq:distribution} and \ref{eq:goldenrule} are the single-particle wave functions $\Psi_{j}(\mathbf{r})$ and their energies $E_{j}$ of the spherical nanoparticle. These quantities are obtained by solving the Schr\"odinger equation for non-interacting electrons in a spherical well potential. For sufficiently large nanoparticles, it has been shown that the effect of electron-electron interactions on hot carrier generation rates is small\cite{nordlander1}. For a given nanoparticle radius, the depth of the potential well is chosen such that the energy of the highest occupied state is equal to experimentally measured work function of the material under consideration (see Table~\ref{tab:parameters}). Note that the work function depends in general on the Miller indices of the surface. For simplicity, we worked with an averaged workfunction when more than one value was available corresponding to a polycrystalline sample. Additional details of our approach to solving the Schr\"odinger equation for the spherical well can be found in the Methods section. In our calculations, we replaced Delta-function in Eq.~\ref{eq:distribution} by a Gaussian with a standard deviation $\sigma$ of $0.05$ eV.

For a given photon energy $\hbar \omega$, integrating Eq.~\ref{eq:distribution} over the hot electron (hole) energy yields the total number of hot electrons (holes) produced per unit time and volume according to
\begin{equation}
N_{e(h)}(\omega)=\int_{-\infty}^{+\infty} N_{e(h)}(E,\omega) dE = \sum_{if} \Gamma_{if}(\omega),
\label{eq:total_number}
\end{equation}
and the power absorbed by the hot carriers is given by
\begin{equation}
P_{hc}(\omega)=\int_{-\infty}^{+\infty} (N_e(E,\omega) + N_h(E,\omega))|E| dE = \sum_{if} \Gamma_{if}(\omega) (|E_i| + |E_f|),
\label{eq:hc_power}
\end{equation}
where all energies are measured with respect to the Fermi level.
Similarly, following \citeauthor{nordlander1}, we define a figure of merit (FoM) as the number of hot electrons above a certain energy $\delta E$ from the Fermi level according to
\begin{equation}
N^{\delta E}_{e}(\omega)=\int_{\delta E}^{+\infty} N_{e}(E,\omega) dE = \sum_{\substack{if\\E_f>\delta E}} \Gamma_{if}(\omega).
\label{eq:fom}
\end{equation}
The FoM of hot holes is obtained by performing the sum with the condition $E_i<-\delta E$.

Finally, the total number of hot carriers produced by a light source of spectral irradiance $S(\omega) = dI(\omega)/d\omega$, where $I(\omega)$ denotes the energy deposited by unit time and area, is given by
\begin{equation}
N=\frac{1}{I_0}\int_0^{+\infty} N_{hc}(\omega) S(\omega) d\omega,
\label{eq:total_number2}
\end{equation}
where $N_{hc}(\omega) = N_e(\omega) + N_h(\omega)$ and $I_0 = 1 \, mW/\mu m^2$ is the intensity of the monochromatic light source used in Eq. \ref{totalpotential}.

We also study optical properties of spherical nanoparticles. If the nanoparticles radius is smaller than the wavelength of light, light absorption dominates over scattering processes \cite{plasmonics2}. Within the quasistatic approximation, the absorption cross section is given by 
\begin{equation}
C_{abs}(\omega) = 4\pi \frac{\omega}{c}n_m \, R^3 \, \Im \left(\frac{\epsilon(\omega)
- \epsilon_m}{\epsilon(\omega) + 2 \epsilon_m} \right),
\label{eq:absorption}
\end{equation}
where $c$ is the speed of light and $n_m$ is the refractive index of the medium.
The power absorbed by the nanoparticle in the quasistatic approximation is
\begin{equation}
P_{qs}(\omega)=I_0 \, C_{abs}(\omega).
\label{eq:qs_power}
\end{equation}

As the radius of the nanoparticle becomes comparable to the wavelength of light, corrections to the quasi-static absorption cross section become important. In particular, Mie theory predicts a red shift of the plasmon energy as the radius increases \cite{size_env_dependence1,plasmonics2,dionne}. The quasi-static approximation also fails for very small nanoparticles when the radius approaches the electron mean free path of the metal \cite{dionne}.

%%%%%%%%%%%%%%%%%%%%%%%%%%%%%%%%%%%%%%%%%%%%%%%%%%%%%%%%%%%%%%%%%%%%%%%%%%%%%%%%%%%%%%%%%%%%%%%%%%%%%%%%%%%%%%%

\begin{figure}[t]
\includegraphics{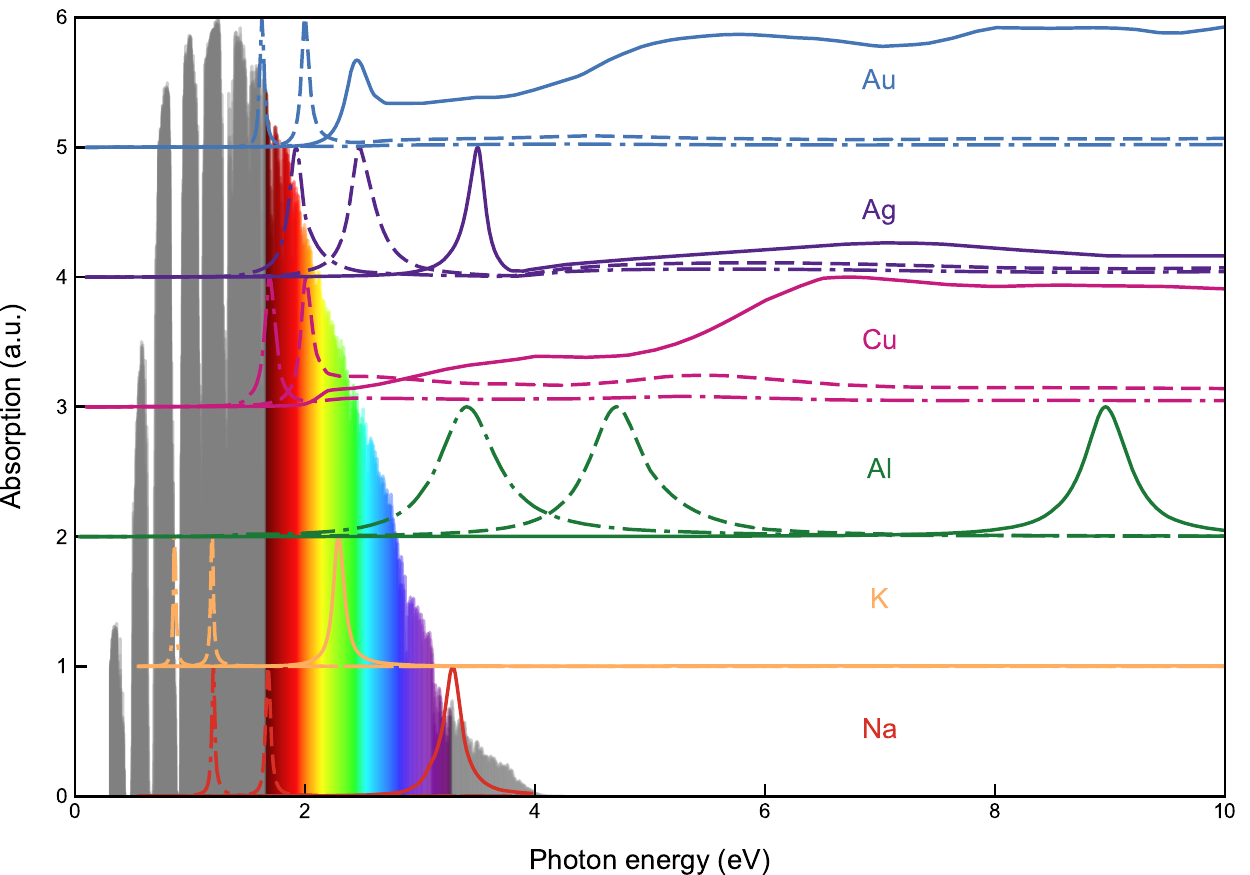}
\caption{Nanoparticle absorption cross section of six plasmonic metals in different environments. Solid, dashed and dash-dotted lines represent
$\epsilon_m = 1$, $\epsilon_m = 5$ and $\epsilon_m = 10$, respectively.
All curves are normalized to their maximum value and are therefore independent of the nanoparticle radius. The shaded curve in the background is the direct plus circumsolar spectral irradiance $S(\omega)$.}
\label{fig:absorption}
\end{figure}

\textbf{Optical properties.}
Figure \ref{fig:absorption} shows the quasi-static absorption cross section for spherical nanoparticles of the six selected plasmonic metals (Au, Ag, Cu, Al, K, Na) as a function of the photon energy and for different dielectric environments. In particular, we carried out calculations for $\epsilon_m=1$ corresponding to a nanoparticle in vacuum, $\epsilon_m=5$ corresponding roughly to a nanoparticle embedded in an organic material (for example, pentacene) and $\epsilon_m=10$ corresponding to a nanoparticle embedded in a semiconductor, such as silicon or GaAs. All curves are offset for clarity and normalized such that the maximum absorption cross section is unity (as a consequence, the results are independent of the nanoparticle radius). The solar spectrum (standard direct plus circumsolar) \cite{solarspectrum} is also drawn as a shaded curve.

We observe a clear difference between simple metals (Al, K and Na) and transition metals (Au, Ag and Cu). The simple metals exhibit a Drude-like absorption spectrum with a single peak corresponding to the localized surface plasmon resonance. The high plasmon energy of Al results from its high conduction electron density. For larger environment dielectric constants, the plasmon peaks shift to lower energies as the additional screening of the surrounding medium reduces the energy required to polarize the nanoparticle. The shift is largest for Al, whose plasmon peak moves from 8.96 eV in vacuum to 3.40 eV for $\epsilon_m=10$ and now overlaps with the solar spectrum. Table~\ref{tab:resonances} summarizes the energies of the localized surface plasmon resonances of the selected metals in different environments.

\begin{table}
\caption{Localized surface plasmon energies (in eV) of spherical nanoparticles in different dielectric environments calculated in the quasistatic approximation.}
\begin{center}
\begin{tabular}{ccccccc}
\hline
$\epsilon_m$ & Na & K & Al & Cu & Ag & Au \\
\hline
1  & 3.28 & 2.29 & 8.96 & 2.15 & 3.50 & 2.40 \\
5  & 1.68 & 1.19 & 4.71 & 2.00 & 2.48 & 2.00 \\
10 & 1.20 & 0.87 & 3.40 & 1.69 & 1.92 & 1.62 \\
\hline
\end{tabular}
\end{center}
\label{tab:resonances}
\end{table}

In contrast, the absorption spectra of the transition-metal nanoparticles exhibit a more complicated shape. For Ag, the spectrum has a strong plasmon peak at 3.50 eV followed by a broad shallow peak at higher energies (for $\epsilon_m=1$). This additional feature results from transitions involving d-bands which -- for Ag -- sit about 4 eV below the Fermi level. For Au and Cu, the d-bands are much closer to the Fermi level (roughly 2.5 eV below it) resulting in a spectrum with a significantly reduced plasmon peak followed by a strong d-band structure. 

Increasing the environment dielectric constant leads to dramatic changes in the absorption spectra of transition-metal nanoparticles. As the plasmon energy is reduced, the d-band features become less important and a Drude-like spectrum is recovered. 

We have compared the calculated absorption spectra with experimental measurements \cite{exptabs1,exptabs2,exptabs3,exptabs5} and found good agreement for the positions of the absorption peaks and also for the shifts induced by a dielectric environment.

\begin{figure}
\includegraphics{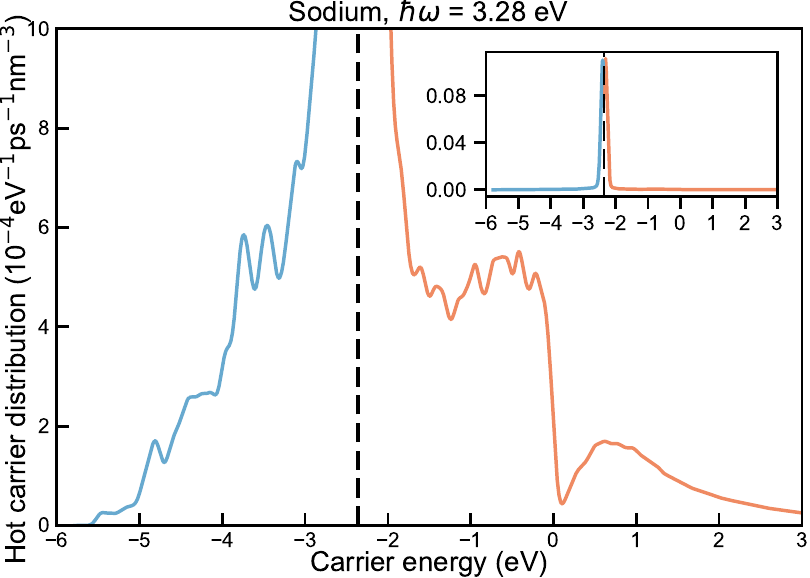}
\caption{Hot electron (red) and hole (blue) distributions for a sodium spherical nanoparticle of 10 nm radius. The photon energy is set to the energy of the localized surface plasmon. The dashed vertical line denotes the Fermi energy. The zero of energy is set to the vacuum level. The hot carrier distributions were rescaled such that the total energy of the hot carriers is equal to the energy absorbed by the nanoparticle.}
\label{fig:sodium}
\end{figure}

%%%%%%%%%%%%%%%%%%%%%%%%%%%%%%%%%%%%%%%%%%%%%%%%%%%%%%%%%%%%%%%%%%%%%%%%%%%%%%%%%%%%%%%%%%%%%%%%%%%%%%%%%%%%%%%
\textbf{Hot carrier energy distribution.} Figure~\ref{fig:sodium} shows the distribution of hot electrons and holes as a function of their energy for a Na nanoparticule with $R=10$~nm.
An environment dielectric constant of $\epsilon_m=1$ was used and the photon energy was set to $\omega_{LSP} = 3.28$~eV, the energy of the localized surface plasmon. We find (see inset) that both hot electron and hole distributions are strongly peaked in the vicinity of the Fermi level $E_F$. At lower (higher) energies, the hole (electron) distribution has a complicated profile with many peaks and decays to zero for energies smaller (larger) than $E_F-\omega_{LSP}$ ($E_F+\omega_{LSP}$).

This behaviour of the hot carrier distributions results from the competition between contributions to the density of final states (Eq.~\ref{eq:finalstates}) from resonant and anti-resonant transitions. Anti-resonant transitions are strong when the transition energy is small and lifetimes are short (they vanish in the limit of infinite lifetimes) and thus give rise to the peaks near the Fermi level. Resonant transitions give significant contributions when lifetimes are long and the transition energy is equal to the photon energy. They give rise to electron and hole distributions that look like shifted copies of each other.

Moreover, the energy-dependence of the carrier lifetimes results in a shift of the resonant peak towards the Fermi level. The magnitude of this shift increases for larger transition energies. This effect thus reduces the number of hot carriers. Promising systems for hot carrier generation should therefore feature weak electron-electron interactions (as those give rise to energy-dependent lifetimes) and low plasmon energies.

Another interesting feature in Fig.~\ref{fig:sodium} is the peak in the hot electron distribution at $\sim$~0.7 eV. These electrons have energies above the vacuum level and can escape from the nanoparticle. Such plasmon-induced ionization processes are possible when the energy of the localized surface plasmon is larger than the work function of the material. The resulting hot electrons can be harvested without the need to extract them from the nanoparticle through transport processes that inevitably lead to hot carrier cooling. This mechanism could therefore be interesting for energy technology applications, such as solar cells or photoelectrochemical devices.   

\begin{figure}
\includegraphics{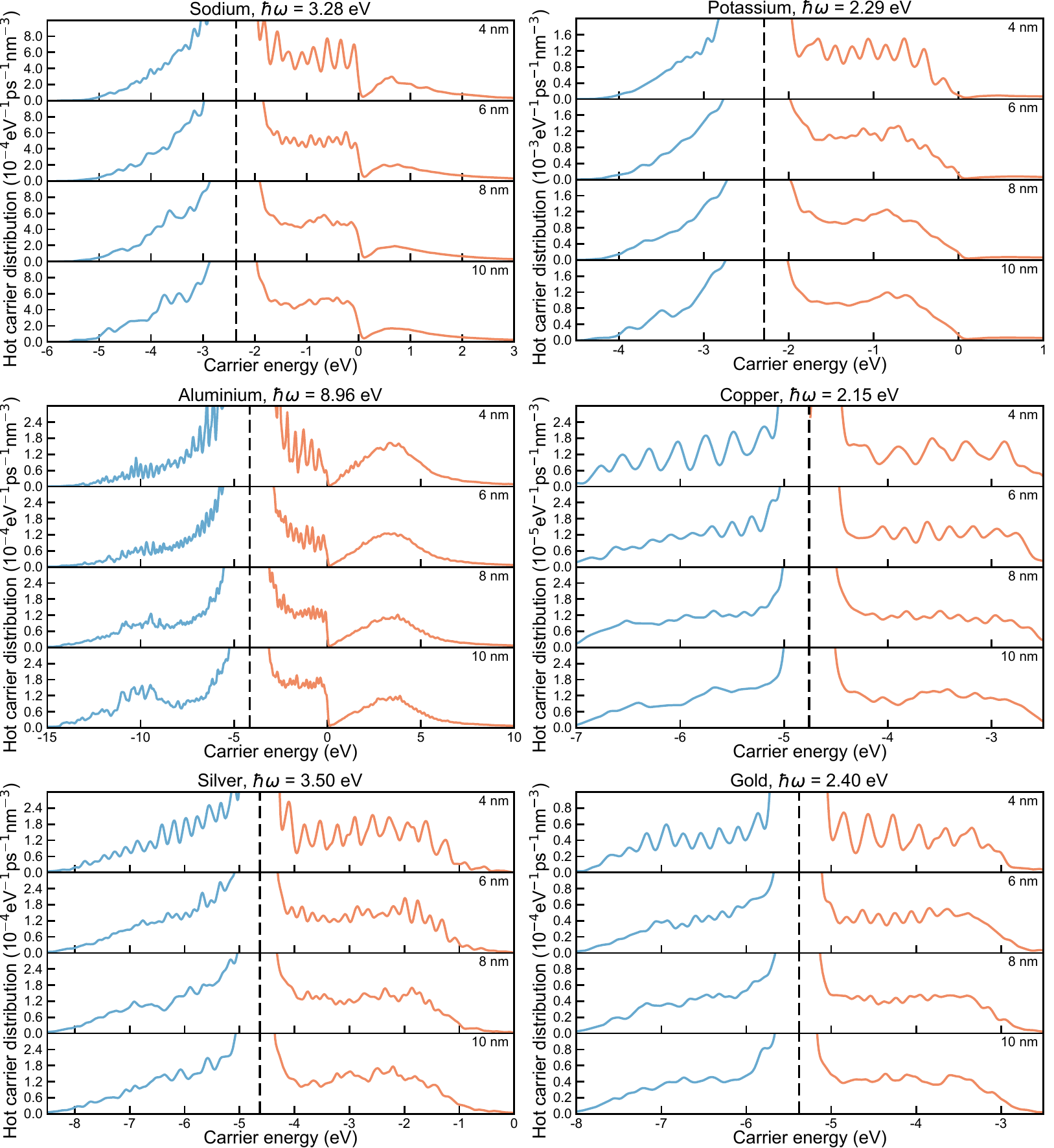}
\caption{Hot electron (red) and hot hole (blue) distributions for spherical nanoparticles of different compositions and sizes. Photon energies are set to the energy of the localized surface plasmon. The dashed vertical line denotes the Fermi energy. The zero of energy is set to the vacuum level. The hot carrier distributions were rescaled such that the total energy of the hot carriers is equal to the energy absorbed by the nanoparticle.}
\label{fig:dist}
\end{figure}

Figure~\ref{fig:dist} shows the distribution of hot electrons and holes
as a function of the hot carrier energy for nanoparticles of different compositions and sizes. Again, an environment dielectric constant of $\epsilon_m=1$ was used and the photon energy was set to the energy of the localized surface plasmon resonance (see Table~\ref{tab:resonances}).

We first discuss the size dependence of the hot carrier distributions. For very small nanoparticles (with radii of 4 nm or less), the distributions exihibit discrete peaks. The electronic structure of such nanoparticles is molecule-like with a discrete set of one-electron levels and the peaks in the hot carrier distribution result from transitions between such levels. For larger nanoparticles, the energy spacing between one-electron levels shrinks and the hot carrier distributions become quasi-continuous. 

Considering the material dependence, we find that all six plasmonic metals exhibit hot carrier distributions with strong peaks near the Fermi energy. Those peaks are largest for Al, which has the highest plasmon energy and the shortest lifetime for carriers near $E_F$, see Fig.~\ref{fig:lifetimes}. Interestingly, because of its high electron density, Al features the longest lifetimes for carriers far from the Fermi level which gives rise to the structure of fine sharp peaks in the hot carrier distributions. The localized surface plasmon energies of simple-metal nanoparticles (Na, K and Al) are similar to or larger than the their workfunctions and as a consequence the hot electron distributions in these materials have peaks near or above the vacuum level. In contrast, the energy of the localized surface plasmon in transition-metal nanoparticles (Au, Ag and Cu) is smaller than the workfunction and the resulting hot carrier distributions are relatively structureless and exhibit no peaks above the vacuum level.

%%%%%%%%%%%%%%%%%%%%%%%%%%%%%%%%%%%%%%%%%%%%%%%%%%%%%%%%%%%%%%%%%%%%%%%%%%%%%%%%%%%%%%%%%%%%%%%%%%%%%%%%%%%%%%

\begin{figure}
\includegraphics{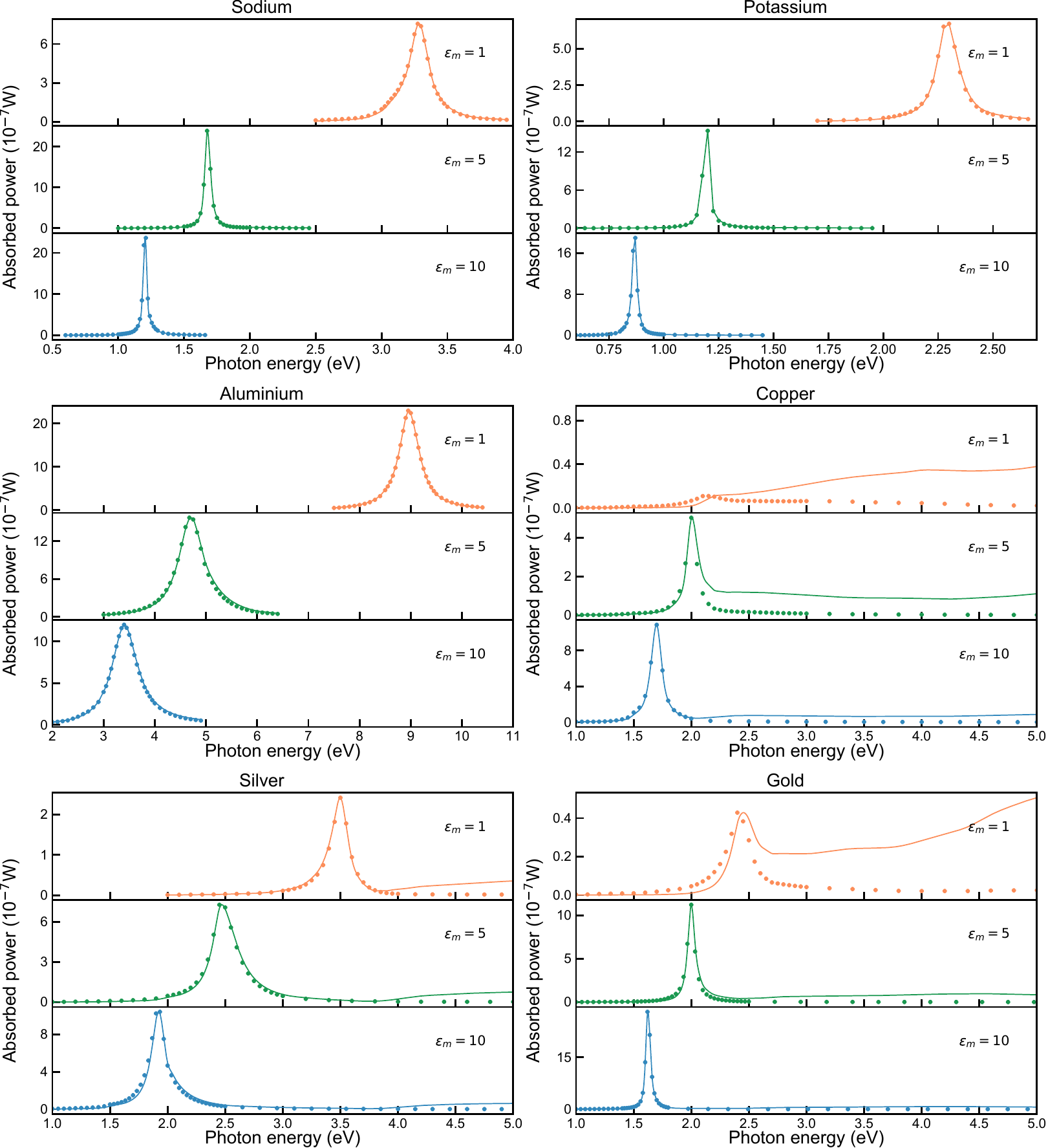}
\caption{Quasistatic absorbed power (solid lines) calculated using Eq.~\ref{eq:qs_power} and the hot carrier absorbed power (dots) calculated using Eq.~\ref{eq:hc_power} for an $R=6$~nm nanoparticle. The hot carrier power has been rescaled to match the maximum value of the quasistatic curves.}
\label{fig:power}
\end{figure}

\textbf{Hot carrier power absorption.} Figure~\ref{fig:power} compares the absorbed power calculated in the quasistatic approximation (Eq.~\ref{eq:qs_power}) with the absorbed power computed from the hot carrier distributions (Eq.~\ref{eq:hc_power}) for an $R=6$~nm nanoparticle made of one of the six selected plasmonic materials in different dielectric media. The illumination intensity is set to 1~mW/$\mu$m$^2$. In general, the two ways of calculating the absorbed power give different results. Two factors contribute to this discrepancy: i) the inability of the spherical well model to describe d-states and ii) finite carrier lifetimes that enable transitions which do not conserve energy (see Eq.~\ref{eq:finalstates}). To distinguish these factors, we rescaled the hot carrier curves in Fig.~\ref{fig:power} such that the height of their first peak is equal to the corresponding feature in the quasistatic curves. 

For the simple metals (Na, K and Al), we find that the total power of the hot carriers $P_{hc}(\omega)$ follows closely the quasistatic absorption $P_{qs}(\omega)$ indicating that the spherical well model accurately describes the electronic structure of these systems. In contrast, there are clear differences between the two curves for transition-metal nanoparticles. For Ag, Au and Cu, $P_{hc}(\omega)$ has a peak at $\omega_{LSP}$ and decreases at higher photon energies, while $P_{qs}(\omega)$ goes through a minimum after the plasmon peak and then increases again. As discussed above, the increase in $P_{qs}(\omega)$ is caused by d-band transitions (recall that we are using experimentally measured bulk dielectric functions to calculate $P_{qs}$ which include d-band transitions as well as other contributions, such as phonon-assisted transitions). The spherical well model does not capture such transitions and the corresponding features in $P_{hc}(\omega)$ are missing.

Note, however, that an increase in the environment dielectric function reduces the relative importance of d-state features in $P_{qs}(\omega)$ for transition-metal nanoparticles and improves the agreement with $P_{hc}(\omega)$. We also observe that higher values of $\epsilon_m$ lead to higher absorption at $\omega_{LSP}$ (except for Al) indicating that there exists an optimal value for the medium dielectric constant which maximises hot carrier generation rates.

\begin{figure}
\includegraphics{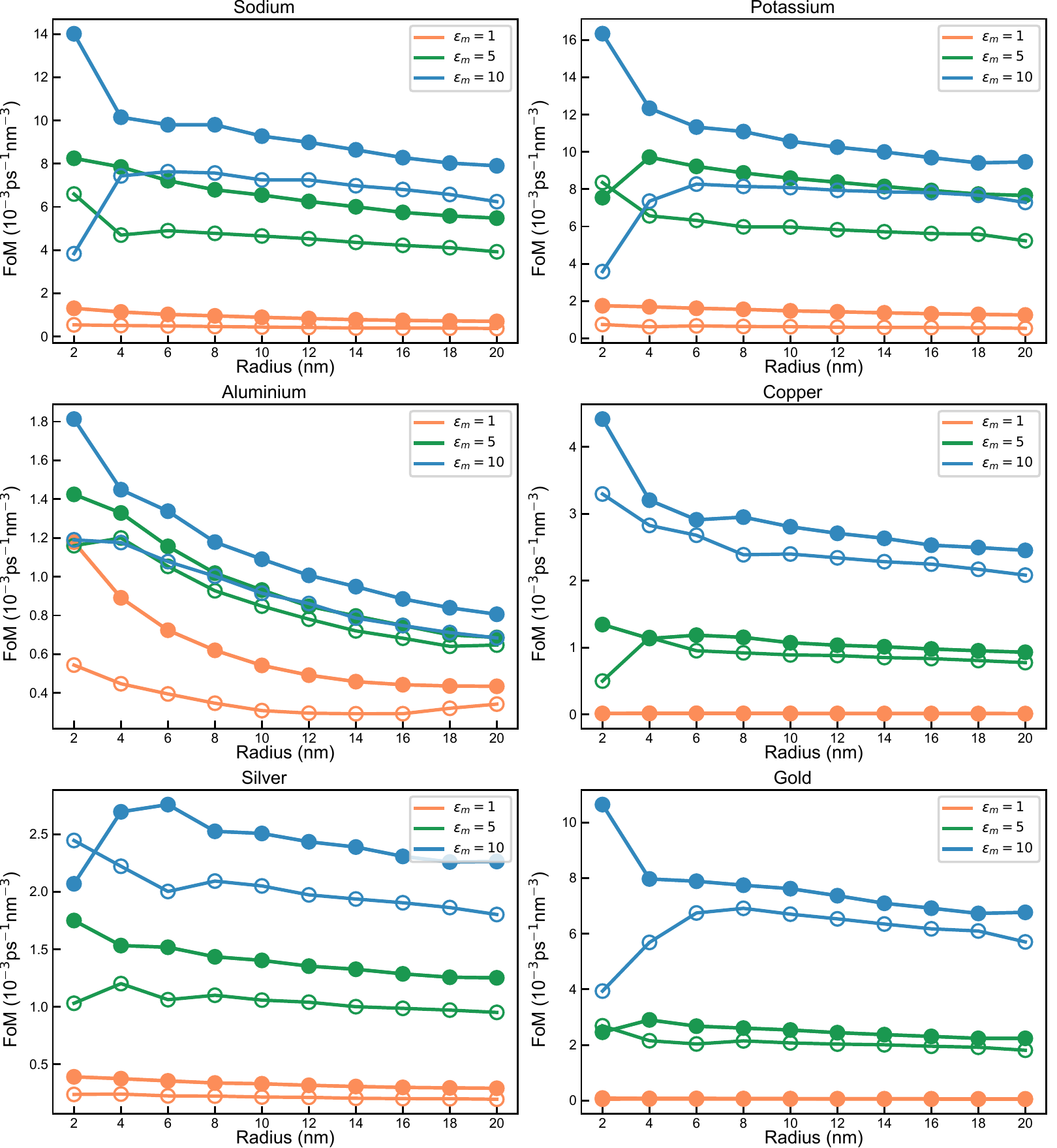}
\caption{Figure of Merit of hot electrons (solid circles) and hot holes (empty circles) as function of nanoparticle radius and environment dielectric constant. $\delta E$ in Eq.~\ref{eq:fom} was set to $0.3 \, \hbar \omega_{LSP}$.}
\label{fig:fom}
\end{figure}

%%%%%%%%%%%%%%%%%%%%%%%%%%%%%%%%%%%%%%%%%%%%%%%%%%%%%%%%%%%%%%%%%%%%%%%%%%%%%%%%%%%%%%%%%%%%%%%%%%%%%%
\textbf{Hot carrier Figure of Merit.} Figure~\ref{fig:fom} shows the FoM (Eq.~\ref{eq:fom}) for nanoparticles of different sizes and compositions in various dielectric environments. We use the energy threshold $\delta E = 0.3 \, \hbar \omega_{LSP}$, where $\omega_{LSP}$ denotes the energy of the localized surface plasmon (see Table~\ref{tab:resonances}).

We observe that the FoM of hot electrons (solid circles) is always larger than the FoM of hot holes (empty circles) (except in some cases for very small nanoparticles). This is due to the fact that electrons can be excited to arbitrarily high energies, while hole energies are restricted by the depth of the potential well. We also find that an increase in the environment dielectric constant $\epsilon_m$ results in an increase in the FoM. Specifically, the FoM of Al nanoparticles doubles as $\epsilon_m$ increases from 1 to 10, while the FoM of Na, K and Ag increases by a factor of 10. For Cu and Au, the FoM even increases by \emph{two orders of magnitude} demonstrating the important role of environment screening for hot carrier properties. The increase in the FoM is caused by two factors: (i) increasing $\epsilon_m$ leads to enhanced absorption at the plasmon resonance (see Fig.~\ref{fig:power}) and (ii) increasing $\epsilon_m$ reduces the plasmon energy and thus increases the relative importance of resonant transitions. 

Figure~\ref{fig:fom} also shows that in general the FoM decreases as the nanoparticle size is increased. For larger nanoparticles, a larger fraction of the excited carriers undergo anti-resonant transitions and have energies closer to the Fermi level. Exceptions to this trend are the FoM of holes in Na, K and Au and the FoM for electrons in Ag for nanoparticles in environments with high dielectric constants. These systems exhibit a maximum in the FoM for radii in the range of 6-8 nms.

\begin{figure}
\includegraphics{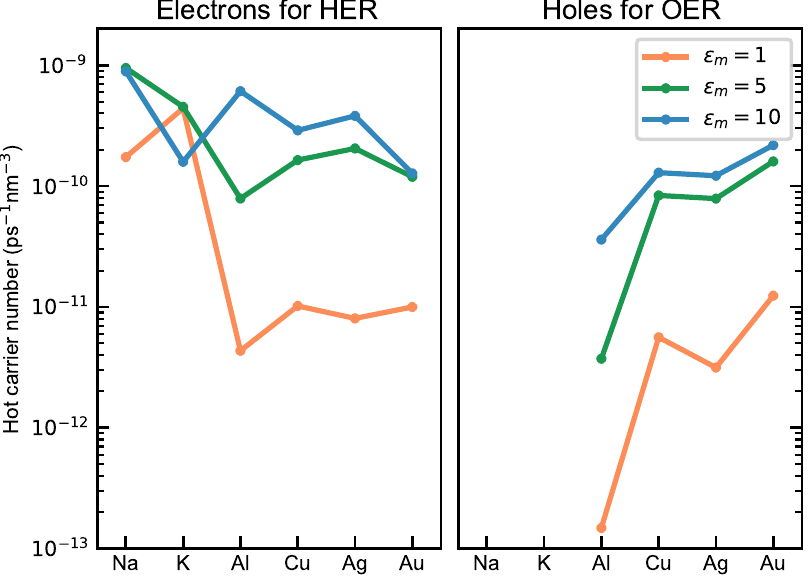}
\caption{Total number of hot electrons and holes produced after illuminating the nanoparticles with sunlight for different environment dielectric functions.
Only electrons above the hydrogen evolution energy (HER) and holes below the oxygen evolution energy (OER) have been considered.}
\label{fig:hydroxy}
\end{figure}

%%%%%%%%%%%%%%%%%%%%%%%%%%%%%%%%%%%%%%%%%%%%%%%%%%%%%%%%%%%%%%%%%%%%%%%%%%%%%%%%%%%%%%%%%%%%%%%%%%%%%%%%%%%%%%%
\textbf{Water splitting.} Plasmonic nanoparticles have been used as catalysts in the photoelectrochemical splitting of water into oxygen and hydrogen gas \cite{watersplitting1,watersplitting2,watersplitting3}. To identify promising candidate systems, we calculate the total number of hot electrons (holes) produced by sunlight (Eq.~\ref{eq:total_number2}) with energies larger (less) than the hydrogen (oxygen) evolution energy $E_{HER}=-4.44$~eV ($E_{OER}=-5.67$~eV) \cite{evolution_energies}. Fig.~\ref{fig:hydroxy} shows our results for $R=6$~nm nanoparticles in different dielectric environments (the radial dependence of water splitting properties can be found in the Supporting Information). We find that simple-metal nanoparticles (most notably, Na and K) are significantly more efficient at producing hot electrons for the hydrogen evolution reaction than transition-metal nanoparticles. This is because the Fermi levels of Na and K sit at a higher energy than in Au, Ag and Cu and therefore the number of hot electrons above the hydrogen evolution energy is larger. As $\epsilon_m$ increases, the total number of hot electrons increases and the difference between simple metals and transition metals becomes smaller. 

%JL: add citation
In contrast, holes for the oxygen evolution reaction are much more efficiently produced in transition-metal nanoparticles,  while the number of hot holes produced in Na and K is extremely small. Again, hot hole production rates are enhanced as $\epsilon_m$ increases. It is important to note that nanoparticles can undergo significant modifications in realistic environments. For example, alkali metals react with aqueous solutions to form dissolved hydroxide ions. Aluminium nanoparticles acquire a surface oxide layer which can act as a tunnel barrier for hot electrons or allow molecules to diffuse to the aluminium surface \cite{zhou2016aluminum}.

\section{Conclusions}
We have studied properties of plasmon-induced hot electrons and holes in spherical nanoparticles and analyzed their dependence on the material, the nanoparticle size and its environment. Our theoretical calculations are based on Fermi's golden rule for the decay of the localized surface plasmon into electron-hole pairs via the Landau damping mechanism. Plasmon properties are described by the quasistatic approximation with experimental bulk dielectric functions and electronic energies and wavefunctions are determined by solving Schr\"odinger's equation for a spherical potential well. We find that hot carrier distributions depend sensitively on carrier lifetimes which we calculate by combining Debye theory for electron-phonon scattering with Fermi liquid theory for electron-electron scattering. An important limitation of the spherical well model is its inability to describe d-states. By comparing the calculated total number of hot carriers with the nanoparticle absorption spectrum, we find that d-states play an important role in transition-metal nanoparticles in weak dielectric environments. However, even for those systems our calculations provide a useful lower bound for hot carrier generation rates.  

For all materials, the hot electron and hole distributions are peaked near the Fermi level because of strong anti-resonant transitions enabled by finite carrier lifetimes. Hot carriers are mostly produced by resonant transitions in materials with long carrier lifetimes and low plasmon energies embedded in environments with large dielectric constants. For simple-metal nanoparticles, plasmon-induced ionization is possible when the energy of the localized surface plasmon is larger than the material's workfunction. We model nanoparticles with diameters up to 40 nm and find that small nanoparticles with diameters less than 16 nm exhibit the highest figure of merit. Among the different studied materials, most hot carriers are produced by Na, K and Au nanoparticles embedded in media with large dielectric constants. Finally, plasmon-induced water splitting was studied. We find that simple metals, in particular Na and K, efficiently generate electrons for the hydrogen evolution reaction while transition-metal nanoparticles produce holes for the oxygen evolution reaction. This suggests that either mixtures of simple-metal and transition-metal nanoparticles or bimetallic nanostructures, such as Janus or core-shell particles, should be used to achieve optimal performance in water splitting devices.

Future work should address the shortcomings of the theoretical approach, in particular the lack of d-states and the empirical description of plasmon properties. To capture d-states within a non-atomistic approach, it is possible to generalize effective mass models to multiple bands with non-parabolic dispersion relations.
%In the limiting case of dispersionless d-bands, we find that hot-carrier generation rates increase significantly (approximately by a factor of 4) when transitions from s- to d-bands are energetically possible \cite{dalforno}.
A parameter-free description of plasmon properties is possible through a direct calculation of the nanoparticle's polarizability within the random-phase approximation or time-dependent density-functional theory\cite{plasmon_breathing,naked_capped_plasmon}.

\begin{acknowledgement}
The authors thank helpful discussions with Peter Nordlander, Ortwin Hess, Vincenzo Giannini and Ravishankar Sundararaman. S.D.F. and J.L. acknowledge support from EPRSC under Grant No. EP/N005244/1 and also from the Thomas Young Centre under Grant No. TYC-101. Via J.L.'s membership of the UK's HEC Materials Chemistry Consortium, which is funded by EPSRC (EP/L000202), this work used the ARCHER UK National Supercomputing Service.
\end{acknowledgement}

%%%%%%%%%%%%%%%%%%%%%%%%%%%%%%%%%%%%%%%%%%%%%%%%%%%%%%%%%%%%%%%%%%%%%%%%%%%%%%%%%%%%%%%%%%%%%%%%%%%
\section{Methods}

We model electrons in spherical metallic nanoparticles as free particles in a spherical potential well given by
\begin{equation}
\label{squarewell}
V(r)=\begin{cases}
V_0, & r \le R, \\
0, & r>R,
\end{cases}
\end{equation}
where $r$ is the distance from the center of the nanoparticle, $V_0$ and $R$ denote the depth of the potential well and the radius 
of the nanoparticle, respectively. Exploiting spherical symmetry, we split the Schr\"odinger equation into an angular and a radial part.
The solutions of the angular part are the spherical harmonics while the solutions of the radial part can be obtained numerically in terms of the spherical Bessel
function of the first and second kind. Specifically, the radial Schr\"odinger equation is given by 
\begin{equation}
\left[ r^2\frac{d^2}{dr^2} + 2r\frac{d}{dr} + \frac{2mr^2}{\hbar^2}\left( E - V(r)\right) -l(l+1)\right] \chi(r) = 0,
\end{equation}
where $E$ is the energy of the electron state, $m$ is the mass of the electron and $l$ is the angular momentum quantum number.
The radial solution $\chi(r)$ has to be continuous at the origin and at infinity, and its first derivative must be continuous at the surface of the nanoparticle.

Bound states have energies $V_0 < E < 0$ and the corresponding radial wavefunctions are of the form
\begin{equation}
\chi(r)=A\begin{cases}
j_l(k_1r), & r \le R, \\
j_l(k_2r)+ i y_l(k_2r), & r > R,
\end{cases}
\end{equation}
where $j_l$ and $y_l$ are the spherical Bessel function of the first and second kind, respectively, and $A$ is a normalisation constant. The wave vectors $k_1=\sqrt{2m/\hbar^2(E-V_0)}$ and $k_2=\sqrt{2m/\hbar^2E}$ depend on
$E$ and must be chosen to guarantee continuity of the solutions and their first derivatives at $r=R$. This condition leads to
\begin{equation}
\label{zeros}
\frac{h_l(k_2 R)}{h_l'(k_2 R)} \frac{j_l'(k_1 R)}{j_l(k_1 R)} - \frac{k_2}{k_1} = 0,
\end{equation}
where $h_l(r)$ is the spherical Hankel function of the first kind defined as $h_l(r)=j_l(r)+ i y_l(r)$.
This equation can be easily solved numerically and the solutions are the energy eigenvalues $E_{nl}$.

Within this model, a spherical nanoparticle is characterized by two parameters: its radius $R$ and the well depth $V_0$, which is determined by the condition that the calculated Fermi energy of the nanoparticle should be equal to the experimentally measured work function of the bulk metal. Specifically, for a given valence electron density and nanoparticle radius, we calculate the number of electrons in the system, solve the Schr\"odinger equation  with an initial starting guess for $V_0$ and determine the Fermi energy. We then vary $V_0$ until the calculated Fermi energy agrees with the measured work function.

Similarly, unbound states have energies $E > 0$ and the corresponding radial wavefunctions are given by
\begin{equation}
\chi(r)=A\begin{cases}
j_l(k_1r), & r \le R, \\
\alpha_{nlR} \, j_l(k_2r) + \beta_{nlR} \, y_l(k_2r), & r > R,
\end{cases}
\end{equation}
where $\alpha$ and $\beta$ are chosen to guarantee continuity of the solutions and its first derivative at $r=R$ and $A$ is a normalization constant. To discretize the continuous spectrum of unbound states, we impose hard wall boundary conditions at a radius $R_{HW} \gg R$.

\begin{suppinfo}
Radial dependence plots of water splitting hot carriers.
\end{suppinfo}

\bibliography{hot_carrier}

\end{document}